\documentclass[fleqn,12pt,twoside]{article}
\usepackage[headings]{espcrc1}

\readRCS
$Id: espcrc1.tex,v 1.2 2004/02/24 11:22:11 spepping Exp $
\usepackage{graphicx}
\usepackage[figuresright]{rotating}

\newcommand{\AmS}{{\protect\the\textfont2
  A\kern-.1667em\lower.5ex\hbox{M}\kern-.125emS}}
\newcommand{\snn}{\sqrt{s_{NN}}}

\newcommand{\pbar}{\overline{p}}
\newcommand{\pbarp}{\pbar+p}

\newcommand{\stc}{\sigma^2_C}
\newcommand{\stcr}{\sigma^2_{C,raw}}

\title{Forward-Backward Multiplicity Correlations in $\snn=200$ GeV Au+Au Collisions}
\author{Peter Steinberg$^1$ 
(for the PHOBOS Collaboration)\\
B.B.Back$^1$,
M.D.Baker$^2$,
M.Ballintijn$^4$,
D.S.Barton$^2$,
R.R.Betts$^6$,
A.A.Bickley$^7$,
R.Bindel$^7$,
A.Budzanowski$^3$,
W.Busza$^4$,
A.Carroll$^2$,
Z.Chai$^2$,
M.P.Decowski$^4$,
E.Garc\'{\i}a$^6$,
T.Gburek$^3$,
N.George$^{1,2}$,
K.Gulbrandsen$^4$,
S.Gushue$^2$,
C.Halliwell$^6$,
J.Hamblen$^8$,
M.Hauer$^2$,
G.A.Heintzelman$^2$,
C.Henderson$^4$,
D.J.Hofman$^6$,
R.S.Hollis$^6$,
R.Ho\l y\'{n}ski$^3$,
B.Holzman$^2$,
A.Iordanova$^6$,
E.Johnson$^8$,
J.L.Kane$^4$,
J.Katzy$^{4,6}$,
N.Khan$^8$,
W.Kucewicz$^6$,
P.Kulinich$^4$,
C.M.Kuo$^5$,
W.T.Lin$^5$,
S.Manly$^8$,
D.McLeod$^6$,
A.C.Mignerey$^7$,
R.Nouicer$^6$,
A.Olszewski$^3$,
R.Pak$^2$,
I.C.Park$^8$,
H.Pernegger$^4$,
C.Reed$^4$,
L.P.Remsberg$^2$,
M.Reuter$^6$,
C.Roland$^4$,
G.Roland$^4$,
L.Rosenberg$^4$,
J.Sagerer$^6$,
P.Sarin$^4$,
P.Sawicki$^3$,
H.Seals$^2$,
I.Sedykh$^2$,
W.Skulski$^8$,
C.E.Smith$^6$,
M.A.Stankiewicz$^2$,
P.Steinberg$^2$,
G.S.F.Stephans$^4$,
A.Sukhanov$^2$,
J.-L.Tang$^5$,
M.B.Tonjes$^7$,
A.Trzupek$^3$,
C.Vale$^4$,
G.J.van~Nieuwenhuizen$^4$,
S.S.Vaurynovich$^4$,
R.Verdier$^4$,
G.I.Veres$^4$,
P.Walters$^8$,
E.Wenger$^4$,
F.L.H.Wolfs$^8$,
B.Wosiek$^3$,
K.Wo\'{z}niak$^3$,
A.H.Wuosmaa$^1$,
B.Wys\l ouch$^4$\\
\small
$^1$~Argonne National Laboratory, Argonne, IL 60439-4843, USA\\
$^2$~Brookhaven National Laboratory, Upton, NY 11973-5000, USA\\
$^3$~Institute of Nuclear Physics PAN, Krak\'{o}w, Poland\\
$^4$~Massachusetts Institute of Technology, Cambridge, MA 02139-4307, USA\\
$^5$~National Central University, Chung-Li, Taiwan\\
$^6$~University of Illinois at Chicago, Chicago, IL 60607-7059, USA\\
$^7$~University of Maryland, College Park, MD 20742, USA\\
$^8$~University of Rochester, Rochester, NY 14627, USA\\
}

\begin{document}

\maketitle

\begin{abstract}
Forward-backward correlations of charged-particle multiplicities in symmetric bins
in pseudorapidity are studied in order to gain insight into the
underlying correlation structure of particle production in Au+Au collisions.
The PHOBOS detector is used to measure integrated multiplicities
in bins defined within $|\eta|<3$, centered at $\eta$ and covering
an interval $\Delta \eta$.  The variance $\stc$ of
a suitably defined forward-backward asymmetry variable
is calculated as a function of
$\eta$, $\Delta\eta$, and centrality.  It is found to be sensitive to short
range correlations, and the concept of ``clustering'' is used to
interpret comparisons to phenomenological models.
\end{abstract}

\section{Introduction}
It is often
presumed that particle correlations are only ``short-range'' in rapidity.
Short range correlations have been observed at all energies \cite{Ansorge:1988fg},
in two-particle correlation measurements.
However, they have never been shown to be the only source of correlation
in multiparticle production.
Indeed, single-particle distributions of inclusive charged particles produced
in heavy ion collisions reveal a distinct ``trapezoidal'' 
structure stretching across
the full rapidity range available in these reactions.  More importantly,
there is no evidence of an extended boost invariance, which would 
imply independent emission from different rapidity regions.  Rather,
two non-trivial effects are visible in the centrality
dependence of particle production, which are apparently long-range
in rapidity.  Integrating over the full phase
space reveals that particle production is linear with the number of
participants~\cite{Back:2003xk}.  
This occurs despite a significant change in the shape
of the pseudorapidity dependence as a function of centrality, 
which happens to be collision-energy independent in the forward 
region~\cite{Back:2002wb}.  
All of this suggests that charged-particle 
production at mid-rapidity is highly correlated
with particle production in the forward region, which
naturally begs the question of the underlying structure of the 
single-particle
distributions.  In these proceedings, we discuss how to 
take first steps in this direction via the study of forward-backward
multiplicity correlations.

\section{Forward-Backward Correlations from Fluctuations}

In this work, particle correlations are studied by the event-by-event
comparison of the integrated multiplicity in a bin defined in the
forward ($\eta>0$) region, called $N_F$, centered at $\eta$ with 
width $\Delta\eta$,
with the multiplicity measured in an identical bin defined 
in the backward hemisphere (called ``$N_B$'').  With these definitions, one
can construct the event-wise observable $C=(N_F-N_B)/\sqrt{N_F+N_B}$,
and measure its variance $\sigma^2_C$ for a set of events with nominally
similar characteristics (e.g. centrality).
If particle sources tend to emit into the forward
{\it or} backward region, such that the partitioning is binomial, this
leads to $\stc=1$, since $\sigma^2(N_F-N_B)=N_F+N_B$.
Short-range correlations arise if the objects emitted into
either hemisphere tend to break into $k$ particles, each of which
stay close in rapidity (e.g. due to isotropic emission).  
If each cluster decays into exactly $k$ particles, then
$C \rightarrow \sqrt{k} C$, so $\stc \rightarrow k \stc$.
Such intrinsic short-range correlations have in fact been measured
in $\pbarp$ and $p+p$ experiments by direct construction of the 2-particle
correlation function in $\eta$~\cite{Ansorge:1988fg,Alpgard:1983xp}.  
By varying $\Delta\eta$ and measuring the ratio
$4\sigma^2_F/\langle N_F \rangle$ as a function of $N_F+N_B$,
they found an ``effective'' cluster multiplicity ($k_{eff}=\langle k \rangle
+ \sigma^2_k / \langle k \rangle$) of approximately 2 charged
particles, which exceeds the value of 1.5 estimated for
a resonance gas~\cite{Stephanov:1999zu}.

\section{Experimental Setup and Analysis Method}

The data analyzed here were taken with the PHOBOS 
detector~\cite{Back:2003sr} during 
Runs 2 and 4, in 2001 and 2004, respectively.  The pseudorapidity 
acceptance was restricted to that of the ``Octagon'' detector,
which is a tube of silicon sensors covering $|\eta|<3$ and
full azimuth except for a region near midrapidity.  To simplify
the analysis for different values of $\eta$ and $\Delta \eta$, 
only the regions of the detector with complete rapidity coverage were kept, 
restricting the total azimuthal acceptance to $\Delta \phi = \pi$,
in four 45-degree wedges~\cite{Wozniak:2004kp}.
The multiplicity in each bin is estimated event-by-event 
by summing up the angle-corrected deposited energy of all
hits and then dividing by the average energy per particle~\cite{Chai:2005}.
Beyond the usual lower threshold to define a hit in the silicon,
an $\eta$-dependent upper bound of the deposited energy per hit
is applied to reduce the effect of slow secondaries on the fluctuations.
The average $\langle C \rangle$, calculated as a function of
$\eta$, centrality, and event vertex, is subtracted event-by-event
to correct for gaps in the Octagon (a process which,
according to simulations, leaves the fluctuations unaffected).

To provide information that can be directly applied to models, a
procedure was developed to estimate and remove the detector 
effects from the raw measured value of $\stc$ ($\stcr$) 
by using Monte Carlo (MC) 
simulations of the PHOBOS apparatus.  The basic idea is to
assume that $\stcr = \stc + \sigma^2_{det}$, where
$\sigma^2_{det}$ is the contribution from detector effects, 
and use the MC to subtract it on average, leaving
behind only the physical correlations.  However, it is found that
there are several sources which contribute differently as a
function of $\eta$ and combine in quadrature to a nearly-constant value
over the pseudorapidity range covered by the Octagon~\cite{Chai:2005}.
These are corrected on average by calculating $\stc$ using a
modified HIJING simulation with all intrinsic correlations
destroyed, which allows a direct estimation of $\sigma^2_{det}$.
There remains a subdominant correlation of $\sigma^2_{det}$ with
$\stc$ which is also removed.  Finally, we correct for using
only half-azimuth acceptance by $\stc \rightarrow 2(\stc-1)+1$, 
a formula obtained using MC simulations, assuming that the
limited acceptance only affects the clustering properties of
particle production.

Systematic uncertainties were calculated by varying several assumptions 
about the estimation of $\stc$ and found to be around
$\Delta \stc \sim 0.1$.  The bin-to-bin variation of the systematic
error calculation was averaged over $\eta$ to reduce fluctuations
in the error determination procedure.

\section{Results}
After correcting for detector effects, the results on $\stc$ can be
directly compared with model calculations based on charged primary particles.
We have focused mainly on HIJING~\cite{Gyulassy:1994ew} 
and AMPT~\cite{Lin:2004en}, which have been used to
describe various features of heavy ion collisions at RHIC energies.

\begin{figure}[t]
\begin{center}
\includegraphics[width=140mm]{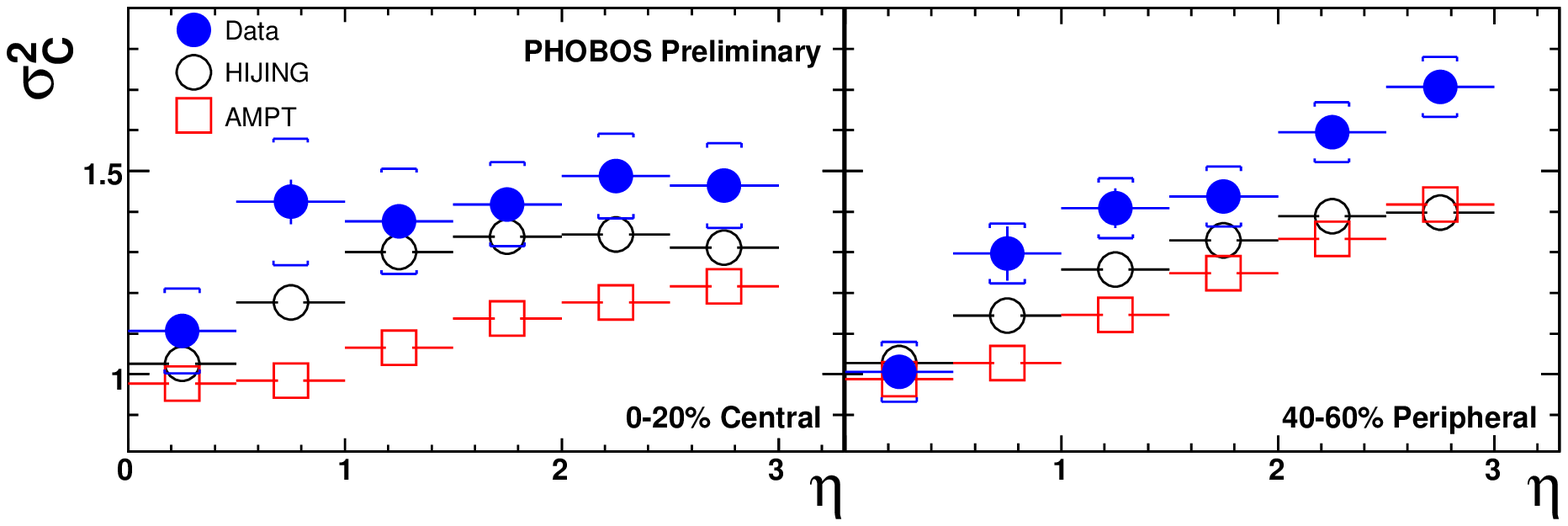}
\includegraphics[width=140mm]{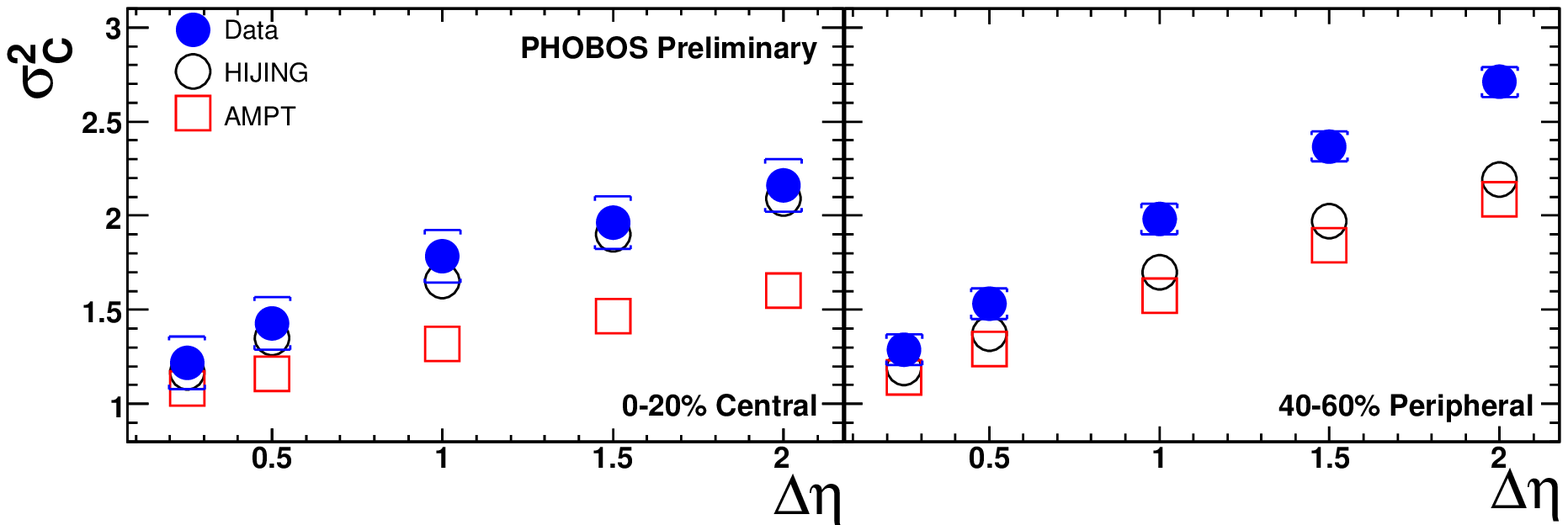}
\vspace*{-1.2cm}
\end{center}
\caption{(upper) $\stc$ for a fixed $\Delta \eta=0.5$ as a function of $\eta$.  (lower) $\stc$ as a function of $\Delta \eta$ for a fixed
bin center at $\eta=2.0$.
\label{fig:etadeta}}
\end{figure}

The first set of results concern the $\eta$ dependence, for
forward and backward bins 
that are $\Delta \eta=0.5$ units wide, as shown in Fig.~\ref{fig:etadeta}.
In this case, HIJING and AMPT already show some interesting
differences.  For peripheral (40-60\%) events, both models 
have a similar magnitude and a monotonically-rising $\eta$ dependence.
Central events show a substantial difference extending over most of
the rapidity range, with AMPT showing a systematically smaller 
value of $\stc$.  This may be due to initially produced clusters
being destroyed by the hadronic rescattering stage.  Also, it is
observed that both data and MC show $\stc \sim 1$ at $\eta = 0$.  
Although this is suggestive of the effect predicted by Jeon et al
\cite{Shi:2005rc},
it can also be explained if clusters produced at $\eta=0$ emit particles into
{\it both} the $N_F$ and $N_B$ side, inducing an ``intrinsic'' long-range
correlation that decreases $\stc$.

The next set of results is the $\Delta \eta$ dependence of $\stc$,
with a fixed $\eta = 2.0$, as shown in Fig. \ref{fig:etadeta}.  
One sees that for both peripheral and
central data, $\stc$ rises monotonically with increasing pseudorapidity
interval.  This can be explained in the context of cluster
emission: by increasing $\Delta \eta$, one
increases the probability of observing more than one particle emitted
from a single cluster in either $N_F$ or $N_B$.  Clearly, the rate of change
with $\Delta \eta$ should reflect the full cluster distribution
(both $\langle k \rangle$ and $\sigma_k$), but these studies have
not yet been done.  It is striking that the peripheral
data has already reached $\stc \sim 3$ for the largest $\Delta \eta$
in peripheral events, while this number is closer to 2 for central
data.  Finally, it is interesting that neither HIJING nor AMPT can
explain both the centrality and $\Delta\eta$ dependence simultaneously,
as HIJING reproduces the central data but has no centrality dependence,
while AMPT is lower than the central data, but also sees a decrease
in the overall magnitude moving from peripheral to central events.

\section{Conclusions}

In conclusion, measurements of forward-backward fluctuations 
provide insight into the structure of long and short range 
correlations in pseudorapidity space.  The new PHOBOS data for
200 GeV Au+Au collisions are now fully corrected for detector 
and background effects, so direct comparisons can be made to
phenomenological models.  We see significant short-range
correlations at all centralities and pseudorapidities, instead of
just at mid-rapidity.  There is a non-trivial centrality and
rapidity dependence of these correlations, in both $\eta$ and
$\Delta\eta$.  Finally, neither HIJING nor AMPT reproduces even
the qualitative features, but the way in which they fail to do so
may well provide information on the underlying physics.  In
particular, more theoretical attention should be paid to the
properties of ``clusters'' required to explain our data.  
Jeon et al have proposed, in Ref.~\cite{Shi:2005rc},
that the formation of a QGP near mid-rapidity
should destroy any sort of cluster structure seen in p+p, and thus
lead to a reduction in $\stc$, beyond the detector-related effects
discussed here.  The data shown here should provide means to 
study such effects, or set upper limits on their occurrence.

\section{Acknowledgements}
%
%
%
%
This work was partially supported by U.S. DOE grants 
DE-AC02-98CH10886,
DE-FG02-93ER40802, 
DE-FC02-94ER40818,  
DE-FG02-94ER40865, 
DE-FG02-99ER41099, and
W-31-109-ENG-38, by U.S. 
NSF grants 9603486, 
0072204,            
and 0245011,        
by Polish KBN grant 1-P03B-062-27(2004-2007),
by NSC of Taiwan Contract NSC 89-2112-M-008-024, and
by Hungarian OTKA grant (F 049823).

\end{document}